\documentclass[useAMS,usenatbib,amsmath]{mn2e}
\usepackage{graphicx}
\usepackage{ulem}
\usepackage{color}
\usepackage{amssymb}
\usepackage{amsmath}
\input{colordvi.tex}

\title{Studying 21cm power spectrum with one-point statistics}
\author[H.Shimabukuro et al.]
  {Hayato Shimabukuro$^1$$^{,2}$,
  Shintaro Yoshiura$^2$, Keitaro Takahashi $^2$, Shuichiro Yokoyama$^3$
  \newauthor % starts a new line in the
   and Kiyotomo Ichiki $^1$ \\
  $^1$Department of Physics, Graduate School of Science, 
Nagoya University, Aichi, 464-8602, Japan\\
  $^2$Department of Physics, Kumamoto University, Kumamoto, Japan\\
  $^3$Department of Physics, Rikkyo University, Tokyo, Japan}

\pagerange{\pageref{firstpage}--\pageref{lastpage}} \pubyear{2002}

\def\LaTeX{L\kern-.36em\raise.3ex\hbox{a}\kern-.15em
    T\kern-.1667em\lower.7ex\hbox{E}\kern-.125emX}

\begin{document}

\label{firstpage}

\maketitle

\begin{abstract}
 The redshifted 21cm line signal from neutral hydrogens is a promising tool to probe the cosmic dawn and the epoch of reionization (EoR). Ongoing and future low-frequency radio experiments are expected to detect its fluctuations, especially through the power spectrum. In this paper, we give a physical interpretation of the time evolution of the power spectrum of the 21cm brightness temperature fluctuations, which can be decomposed into dark matter density, spin temperature and neutral fraction of hydrogen fluctuations. From the one-point statistics of the fluctuations, such as variance and skewness, we find that the peaks and dips in the time evolution are deeply related to X-ray heating of the intergalactic gas, which controls the spin temperature. We suggest the skewness of the brightness temperature distribution is a key observable to identify the onset of X-ray heating.
\end{abstract}

RUP-14-18

\section{Introduction}

At the early epoch of the universe ($z\sim$ 1100), proton couples with free electron, which is called ``recombination'' and neutral hydrogen is formed.  After recombination, the universe keeps neutral state until first object is formed.  We call this epoch ``dark age''. As the universe is evolved, matter fluctuations grow gravitationally and dark matter is collapsed at over-dense region. Gravitational potential by collapsed objects attracts baryonic objects around dark matter halos and these baryonic objects form primordial stars \cite{yos}. As a result of the formation of first stars, the period called ``cosmic dawn'' in which astrophysical processes play an important role begins \cite{Fialkov:2013uwm,Visbal:2012aw,2011A&A...527A..93S}.  After the first objects are formed, the spin temperature of neutral hydrogen couples to the color temperature of Lyman alpha (Ly-$\alpha$) radiation emitted from the first objects. This effect is called ``Wouthuysen Field (WF) effect''\cite{wou}. Typically,  Ly-$\alpha$ radiation field temperature is nearly equal to the kinetic temperature of intergalactic medium (IGM) because Ly-$\alpha$ scattering rate is very large and the frequent scattering leads to a large number of Ly-$\alpha$ photons which brings Ly-$\alpha$ radiation field coupling to the gas which property is determined by the kinetic temperature \cite{fur}.  Therefore, the spin temperature comes to couple to the kinetic temperature.  After the WF effect becomes effective, next astrophysical process which affects the thermal history of IGM is X-ray heating \cite{Pritchard:2006sq}. It is expected that cold IGM is heated due to the X-ray photons emitted from X-ray binaries \cite{Fialkov:2014kta} and gas kinetic temperature increases dramatically \cite{Mesinger:2012ys, Christian:2013gma}.  As the heating of IGM progresses, thermal phase of IGM is changed and neutral hydrogen starts to be ionized by UV radiation from early galaxies \cite{Loeb:2000fc}. This epoch is called ``Epoch of Reionization (EoR)'' \cite{Fan:2006dp}.

The redshifted 21cm line from neutral hydrogen is sensitive to the thermal and ionized states of gas as well as the first objects in the early universe and it has been expected to give us valuable information about the thermal history of intergalactic medium (IGM) \cite{fur}.  

One of the statistical methods to probe the IGM state is the power spectrum of brightness temperature (21cm power spectrum) \cite{fur,Pritchard:2006sq,2008ApJ...689....1S,Baek:2010cm,Mesinger:2013nua,2014ApJ...782...66P}.  Brightness temperature is a fundamental quantity of the 21cm signal, which is the spin temperature offsetting from the CMB temperature $T_{\gamma}$ along the line of sight (LoS) at observed frequency $\nu$. This can be written as \cite{fur}
\begin{align}
\delta T_{b}(\nu) &= \frac{T_{{\rm S}}-T_{\gamma}}{1+z}(1-e^{-\tau_{\nu_{0}}}) \nonumber \\
                  &\sim 27x_{{\rm H}}(1+\delta_{m})\bigg(\frac{H}{dv_{r}/dr+H}\bigg)\bigg(1-\frac{T_{\gamma}}{T_{{\rm S}}}\bigg)  \nonumber\\
                 &\quad \times \bigg(\frac{1+z}{10}\frac{0.15}{\Omega_{m}h^{2}}\bigg)^{1/2}\bigg(\frac{\Omega_{b}h^{2}}{0.023}\bigg) [{\rm mK}].
\label{eq:brightness}
\end{align}
Here, $T_{{\rm S}}$ is gas spin temperature, $\tau_{\nu_{0}}$ is the optical depth at the 21cm rest frame frequency $\nu_{0}$=1420.4${\rm MHz}$, $x_{{\rm H}}$ is neutral fraction of the gas, $\delta_{m}({\bold x},z) \equiv\rho/\bar{\rho} -1$ is the evolved matter overdensity, $H(z$) is the Hubble parameter and $dv_{r}/dr$ is the comving gradient of the line of sight component of the comving velocity of the gas. All quantities are evaluated at redshift $z=\nu_{0}/\nu-1$.  

On-going radio interferometers such as Murchison Wide Field Array \cite{Tingay:2012ps}, Low Frequency Array \cite{Rottgering:2003jh} and Probing the Epoch of Reionization \cite{Pober:2014aca} have been operated as the ``prototype" of future high-sensitivity experiments. Although the sensitivities of the on-going experiments are not enough to obtain the images of the redshifted 21cm line from the early universe, the power spectrum of the 21cm signal would be detected \cite{Mesinger:2013nua}. In particular, it is expected that the 21cm power spectrum for each redshift ($z = 10 \sim 30$) can be measured by Square Kilometre Array with high accuracy \cite{Carilli:2014vha}.

There have been numerous works about the 21cm power spectrum in both theoretical and observational aspects \cite{Mao:2008ug, Loeb:2008hg, Dillon:2013rfa, Pober:2014aca}. In particular, most of such works have focused on the 21cm signal from the epoch of Reionization (EOR) at $z \sim 6 - 10$. In this epoch, the 21cm power spectrum is expected to trace the spatial distribution of the neutral fraction of gas $x_{{\rm H}}$ and we can ignore the fluctuation of the spin temperature because $T_\gamma / T_{\rm S} \ll 1$.
On the other hand, the 21cm signal from $z \gtrsim 10 - 15$, at when in the universe the first astrophysical objects have started to be formed, has not been investigated deeply~\cite{Ghara:2014yfa}. This epoch is called as ``cosmic dawn" and it should be also important to investigate the 21cm signal from the IGM at this epoch because the state of IGM at this epoch has to be crucial for the formation of the first astrophysical objects. At this epoch, since the reionization process has not proceeded sufficiently, the neutral fraction of gas $x_{{\rm H}}$ should be almost unity and the spatial distribution of the brightness temperature does not trace that of the neutral fraction. Instead of this, the 21cm power spectrum from this epoch should have the information about not only the matter density field but also the spin temperature related with the state of IGM.

In this work, we focus on the 21cm signal from the epoch of ``cosmic dawn", that is $z \gtrsim 15$, and investigate what we can learn from the signal by using a public code called 21cmFAST \cite{Mesinger:2007pd,2011MNRAS.411..955M}. This code is based on a semi-analytic model of star/galaxy formation and reionization.
First, in order to find how physical processes are related to the observed 21cm signal at $z \gtrsim 15$, we decompose the 21cm power spectrum into three components composed by fluctuations of the matter density field, those of the neutral fraction and those of  the spin temperature, and discuss the redshift evolution of the power spectrum.

Next, we further focus on the distributions, variances and skewnesses of the brightness temperature and three decomposed components, in order to understand the physical meaning of the behavior of the 21cm power spectrum more deeply.  Although there are several studies which discuss the one-point statistics of the brightness temperature \cite{Harker:2008pz, Watkinson:2013fea}, most of such works have focused on the signal from EoR, that is, $z \lesssim 10$, to investigate its dependence on the reionization process, and for the signal from $z \gtrsim 15$ the one-point statistics are not sufficiently discussed.

This paper is organized as follows. In section 2, first we show the 21cm power spectrum and the decomposition of it into each component and discuss the redshift dependence of the power spectrum. In section 3, we investigate distribution and one-point statistics such as variance and skewness of the spin temperature which is a dominant component of the 21cm power spectrum at higher redshift ($z \gtrsim 15$), and also we investigate its dependence on the X-ray heating efficiency. In section 4, based on the discussion about the one-point statistics of the spin temperature given in section 3, we investigate the one-point statistics of observed brightness temperature. In section 5, we give a summary and conclusion.

Unless stated otherwise, we quote all quantities in comoving units.  We employ the best fit values of the standard cosmological parameters obtained in \cite{Komatsu:2010fb}.

\section{evolution of power spectrum}

In this section, we calculate the 21cm power spectrum and its decomposition into spectra of the matter density field, the neutral fraction and the spin temperature. In particular, we focus on the redshift evolution of the power spectra.

First, we define the power spectrum of brightness temperature as,
\begin{equation}
\langle \delta_{21}({\bold k}) \delta_{21}({\bold k^{'}})\rangle
= (2\pi)^3 \delta({\bold k}+{\bold k^{'}}) P_{21}({\bold k}),
\label{eq:ps_def}
\end{equation}
where $\delta_{21}({\bold k}) \equiv \delta T_b({\bold k}) / \langle \delta T_b\rangle-1$ and $\langle \delta T_b\rangle$ is the mean brightness temperature obtained from brightness temperature map. As we saw in Eq.(\ref{eq:brightness}), the fluctuation of brightness temperature is contributed from those in the matter density, spin temperature, neutral fraction and peculiar velocity. Here we define $\delta_{\rm H}$ and $\delta_\eta$ such that
$x_{\rm H} = \overline{x}_{\rm H} (1 + \delta_{\rm H})$ and
$\eta = \overline{\eta} (1 + \delta_\eta)$ where $\eta = 1-T_\gamma/T_{\rm S}$, $\overline{x}_{{\rm H}}, \overline{\eta}$ are volume average of $x_{{\rm H}}, \eta$
and the power spectra of $\delta_{m}, \delta_{\rm H}$ and $\delta_\eta$ as,
\begin{eqnarray}
&&
\langle \delta_{m}({\bold k}) \delta_{m}({\bold k^{'}})\rangle
= (2\pi)^3 \delta({\bold k}+{\bold k^{'}}) P_{m}({\bold k}),
\label{eq:ps_delta} \\
&&
\langle \delta_{\rm H}({\bold k}) \delta_{\rm H}({\bold k^{'}})\rangle
= (2\pi)^3 \delta({\bold k}+{\bold k^{'}}) P_{x_{\rm H}}({\bold k}), 
\label{eq:ps_H} \\
&&
\langle \delta_{\eta}({\bold k}) \delta_{\eta}({\bold k^{'}})\rangle
= (2\pi)^3 \delta({\bold k}+{\bold k^{'}}) P_{\eta}({\bold k}),
\label{eq:ps_eta}
\end{eqnarray}
Here, we introduce a new parameter $\eta$ in order to describe the fluctuation of the spin temperature $T_{{\rm S}}$.
By making use of  this parameter, the brightness temperature can be related with the contribution
of the spin temperature linearly. In the case where the fluctuation of the spin temperature $\delta_{T_{{\rm S}}} (= (T_{{\rm S}} - \bar{T}_{{\rm S}}) / \bar{T}_{{\rm S}})$ is small,
$\delta_\eta$ and $\delta_{T_{{\rm S}}}$ are related as

\begin{eqnarray}
\delta_{\eta} \simeq \frac{T_\gamma / \bar{T}_{{\rm S}}}{1 - T_\gamma / \bar{T}_{{\rm S}}} \delta_{T_{{\rm S}}} .
\end{eqnarray}
 
The cross correlations such as $P_{x_{\rm H}\eta}$ are defined in a similar way. Then, from Eq.(\ref{eq:brightness}), we see that the power spectrum of brightness temperature can be decomposed into a sum of the auto-correlations and cross-correlations:
\begin{equation}
P_{21}
= (\overline{\delta T_{b}})^2
  [ P_{m} + P_{x_{\rm H}} + P_{\eta} + P_{x_{\rm H}\eta}
    + P_{x_{\rm H}m} + P_{m\eta} ],
\label{eq:ps_higher}
\end{equation}
where, 
\begin{equation}
\overline{\delta T_{b}} = 27 \overline{x_{\rm H}}\ \overline{\eta} [(1+z)/10]^{1/2} (0.15/\Omega_m h^2)^{1/2} (\Omega_b h^2/0.023) \nonumber.  
\end{equation}
We neglected higher-order terms and we have checked that the higher-order contributions are at most $30\%$ of the total power spectrum at higher redshifts we are interested in here, and do not affect the qualitative feature of the redshift evolution. 
Therefore, the contribution can be safely neglected for the purpose of this article.
However, for more precise quantitative discussion, 
such higher-order contributions have to be evaluated precisely.
We plan to investigate the higher-order contribution and non-linearity in the future.

To compute the power spectra, we use a public code called 21cmFAST \cite{Mesinger:2007pd,2011MNRAS.411..955M}. This code is based on a semi-analytic model of star/galaxy formation and reionization, and outputs maps of matter density, velocity, spin temperature, ionized fraction and brightness temperature at the designated redshifts. We performed simulations in a $(200 {\rm Mpc})^3$ comoving box with $300^3$ grids, which corresponds to 0.66 {\rm Mpc} resolution or $\sim$ 3 arcmin at 80 {\rm MHz} (${\it z}$=17) (we show some telescope specifications in table.\ref{tb:spec}), from $z = 200$ to $z = 8$ adopting the following parameter set, $(\zeta, \zeta_{X}, T_{\rm vir}, R_{\rm mfp}) = (31.5, 10^{56}/M_{\odot}, 10^4~{\rm K}, 30~{\rm Mpc})$. Here, $\zeta$ is the ionizing efficiency, $\zeta_{X}$ is the number of X-ray photons emitted by source per solar mass, $T_{\rm vir}$ is the minimum virial temperature of halos which produce ionizing photons, and $R_{\rm mfp}$ is the mean free path of ionizing photons through the IGM. In our calculation, we also ignore, for simplicity, the gradient of peculiar velocity whose contribution to the brightness temperature is relatively small (a few \%) \cite{Ghara:2014yfa}.

\begin{table*}
\centering
\caption{Resolution at 80 MHz ($z \sim 17$) (Dewdney et al 2013). }
\scalebox{1.5}{
	\begin{tabular}{|l|c|r|r|} \hline
	telescope & maximum baseline & spacial resolution & angle resolution\\ \hline \hline
	LOFAR & $\sim$ 1500 m & $\sim$ 13 Mpc  & $\sim$ 4 arcmin\\ \hline
	MWA   & $\sim$  750 m & $\sim$ 27 Mpc & $\sim$ 9 arcmin \\ \hline
	SKA1  & $\sim$ 2000 m & $\sim$ 10 Mpc & $\sim$ 3 arcmin \\ \hline
	SKA2  & $\sim$ 5000 m & $\sim$  4 Mpc & $\sim$1 arcmin\\ \hline
	\end{tabular}
	}
	\label{tb:spec}
\end{table*}

\begin{figure}
\centering 
\includegraphics[width=0.9\hsize]{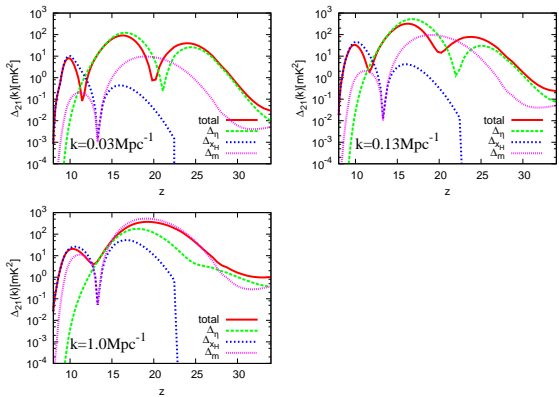}
\caption{Total and decomposed 21cm power spectra as functions of the redshift for $k = 0.03~{\rm Mpc}^{-1}$ (left\ top), $0.13~{\rm Mpc}^{-1}$ (right\ top) and $1.0~{\rm Mpc}^{-1}$ (left\ bottom). 
}
\label{fig:ps_z}
\end{figure}

In Fig. \ref{fig:ps_z}, we plot the total and decomposed power spectra, $\Delta_{i} = k^3 P_i / 2 \pi^2$ with $i = 21, x_{\rm H}, m$ and $\eta$, as functions of redshift for $k = 0.03, 0.13, 1.0~{\rm Mpc}^{-1}$. We can see several characteristic peaks in the total power spectrum: three peaks for $k = 0.03, 0.13~{\rm Mpc}^{-1}$ and two peaks for $k=1.0 {\rm Mpc}^{-1}$. These were already found in the previous works and it was suggested that they are, from high-z one to low-z one, induced by the WF effect, X-ray heating and reionization, respectively \cite{Pritchard:2006sq}. Below we give a more detailed interpretation considering contributions from the fluctuations in neutral fraction, matter density and spin temperature.

First, the fluctuation in neutral fraction appears when reionization begins but power spectrum, $(\overline{\delta T_b})^2 P_{x_{\rm H}}$, is subdominant at high redshifts ($z \gtrsim 15$). It becomes dominant as reionization proceeds and forms the low-z peak at $z \approx 10$. A dip at $z \approx 14$ corresponds to the redshift when the average spin temperature becomes equal to the CMB temperature so that the average brightness temperature, $\overline{\delta T_b}$, vanishes. This dip is also seen in the contribution of matter fluctuations, which is important at smaller scales.

On the other hand, the power spectrum contributed from the spin-temperature fluctuations is negligible at low redshifts ($z \lesssim 10$). This is because at these redshifts the spin temperature is much higher than the CMB temperature and the factor $\eta = 1 - T_\gamma/T_{\rm S}$ is almost unity independent of the value of $T_{\rm S}$. The spin-temperature fluctuations are important at higher redshifts ($z \gtrsim 13$) especially at larger scales and this contribution forms the two peaks at $z \approx 16$ and $24$. However, there are slight deviations in the peak positions between the brightness temperature and spin temperature due to the presence of matter fluctuations.

Thus, the low-z peak and the other two peaks are contributed from the neutral fraction and spin temperature fluctuations, respectively, while the high-z peak at small scales can not be seen due to the contribution from matter fluctuations. In other words, the evolution of spin temperature, which reflects the formation rate and properties of the first-generation stars, can be directly probed by measuring the power spectrum of brightness temperature at $z \gtrsim 15$. In the next section, we focus on the understanding of spin-temperature fluctuations at this epoch considering one-point statistics.

\section{one-point statistics of spin temperature}

In this section, we study the probability distribution function (PDF) of spin temperature and one-point statistics such as variance and skewness. In practice, we focus on $\eta = 1 - T_{\gamma}/T_{\rm S}$, rather than $T_{\rm S}$ itself, because it is linearly related to the brightness temperature. However, because $\eta$ is a monotonic function of $T_{\rm S}$, physical interpretation is relatively straightforward. Recently, one-point statistics of brightness temperature during reionization was investigated assuming $T_{\rm S} \gg T_\gamma$ \cite{2008MNRAS.384.1069B,Harker:2008pz,Watkinson:2013fea}. This condition is not valid in our context.

\begin{figure}
\centering 
\includegraphics[width=1.0\hsize]{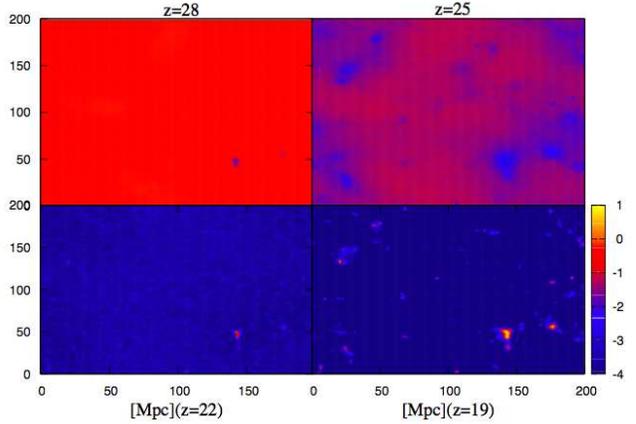}
\caption{The map of $1 - T_\gamma / T_{\rm S}$ at $z = 28$ (left top), $25$ (right top), $22$ (left bottom), $19$ (right bottom). We can see that the spatial averaged value of $\eta$ decreases from $z = 28$ to $19$.
A spatial distribution of $\eta$ can be also seen in each panel. }
\label{fig:Ts_map}
\end{figure}

Obtained maps of $\eta$ from 21cmFAST for the redshifts $z=28, 25, 22,$ and $19$ are shown in Fig.~\ref{fig:Ts_map}. 
From these maps, we evaluate the PDF and also the variance and skewness of $\eta$.
The variance and skewness of a variable $X$ are defined as,
\begin{eqnarray}
&& \sigma^2 = \frac{1}{N} \sum_{i=1}^{N} \big[ X - \overline{X} \big]^2
\label{eq:variance} \\
&& \gamma = \frac{1}{N \sigma^3} \sum_{i=1}^{N} \big[ X - \overline{X} \big]^3,
\label{eq:skewness}
\end{eqnarray}
where $N$ is the number of pixels of the maps. Note that the skewness is negative (positive) when the tail of the distribution relatively extends toward low (high) values of $X$.

\begin{figure}
%\centering \includegraphics[width=0.4\hsize, bb=0 0 350 200]{Tcmb_Ts_combine2.eps}
\centering \includegraphics[width=1.0\hsize]{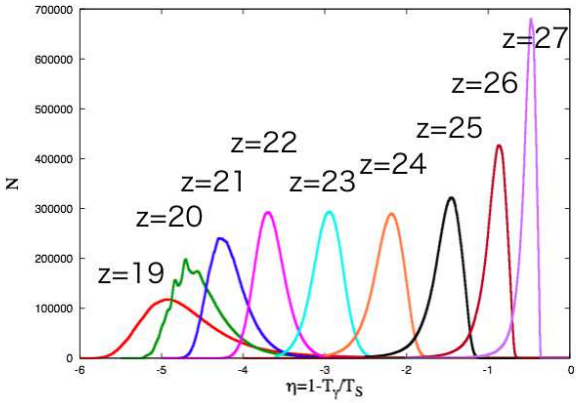}
\caption{PDF of $1-T_\gamma / T_{\rm S}$ for $z = 19-27$ obtained from the map of the spin temperature shown in Fig. \ref{fig:Ts_map}.}
\label{fig:Ts_Tcmb_compare2}
\end{figure}

\begin{figure}
\centering 
\includegraphics[width=1\hsize]{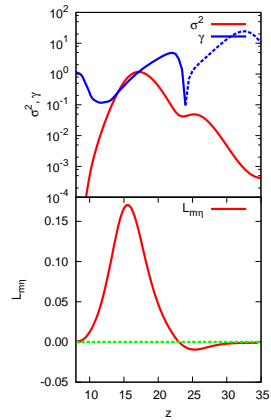}
\caption{[top] Variance and skewness of $1 - T_\gamma / T_{\rm S}$ as functions of redshift. 
For the skewness (denoted as a blue line), the dashed part corresponds to the negative skewness ($\gamma < 0$).
[bottom] Cross correlation between $\delta_m$ and $\eta$ as a function of redshift. Green dashed line corresponds
to $L_{m \eta} = 0$.}
\label{fig:variance_skewness_Ts_Tcmb}
\end{figure}

We plot the PDF of $\eta$ for $z = 19-27$ in Fig.\ref{fig:Ts_Tcmb_compare2}. First of all, the average value of $\eta$ decreases as redshift decreases because the spin temperature strongly couples to the kinetic temperature, which decreases as $(1+z)^2$ due to the adiabatic cooling of gas, via the WF effect at this epoch. Next, we see the shape of the PDF is also changing and, in particular, the direction of the longer tail changes at $z \approx 23$ (cyan dot-dashed line). This is also confirmed in the top of Fig. \ref{fig:variance_skewness_Ts_Tcmb}, where we plot the time evolution of the variance and skewness of the PDF. Actually, the sign of the skewness changes at the same redshift from negative to positive. Here it is important to note that the variance has a local minimum at almost the same timing. Because the variance is the integration of the power spectrum with respect to the wavenumber, this local minimum corresponds to the dip in the contribution of $\eta$ in Fig. \ref{fig:ps_z}.

The above behavior can be understood by considering the X-ray heating of the gas. As we can see in Fig. \ref{fig:Ts_map}, at higher redshifts ($z \gtrsim 25$) the spin temperature in the neighborhood of stars approaches to the kinetic temperature due to the WF effect and becomes lower than the average, and consequently its PDF has a tail toward lower temperature. Then, as X-ray heating becomes effective, the spin temperature increases in the neighborhood of stars and the tail goes toward higher-value side. At the transition time, the tail becomes shortest and consequently the variance has a local minimum there.

The bottom of Fig. \ref{fig:variance_skewness_Ts_Tcmb}, the cross correlation between the matter and spin temperature fluctuations, strongly supports the above interpretation. Here we evaluate the cross correlation from the cross power spectrum as
\begin{eqnarray}
L_{m \eta } := \int {d^3 k \over (2 \pi )^3} P_{m\eta}(k).
\end{eqnarray}
At higher redshifts ($z \gtrsim 23$), the correlation is negative, that is, high-density regions have lower spin temperature. This would be due to the WF effect. On the other hand, at lower redshifts ($z \lesssim 23$), as is expected from the interpretation that X-ray heating is effective here, the correlation is positive. It is seen that the cross correlation changes the sign at the same redshift as the skewness.

\begin{figure}
\centering 
\includegraphics[width=1.0\hsize]{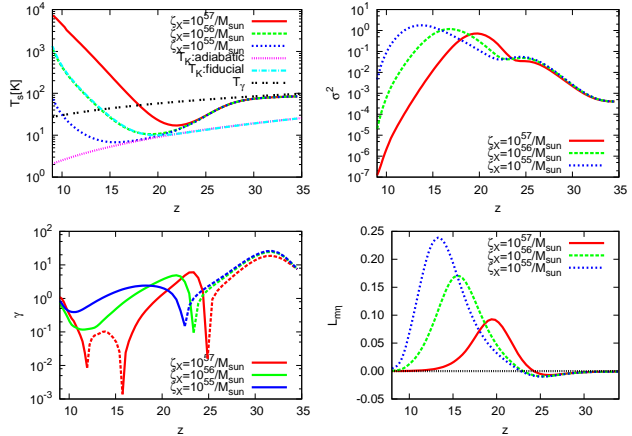}
\caption{[left top] Spin temperature evolution for $\zeta_X = 10^{57}~{\rm M}_{\odot}^{-1}$ (red), $10^{56}~{\rm M}_{\odot}^{-1}$ (green) and $10^{55}~{\rm M}_{\odot}^{-1}$ (blue).  We also plot kinetic temperature in the absence of X-ray heating (pink) and in the presence of X-ray heating with $\zeta_X = 10^{56}~{\rm M}_{\odot}^{-1}$ (light blue) and CMB temperature (black).  [right top and left bottom] Evolution of variance and skewness of the probability distribution function of $1 - T_\gamma / T_{\rm S}$ for $\zeta_X = 10^{57}~M_{\odot}^{-1}$ (red), $10^{56}~M_{\odot}^{-1}$ (green) and $10^{55}~M_{\odot}^{-1}$ (blue). [right bottom] Cross correlation between $\delta_m$ and $\delta_\eta$ for $\zeta_X = 10^{57}~M_{\odot}^{-1}$ (red), $10^{56}~M_{\odot}^{-1}$ (green) and $10^{55}~M_{\odot}^{-1}$ (blue).}
\label{fig:model}
\end{figure}

\begin{figure}
\centering 
\includegraphics[width=1.0\hsize]{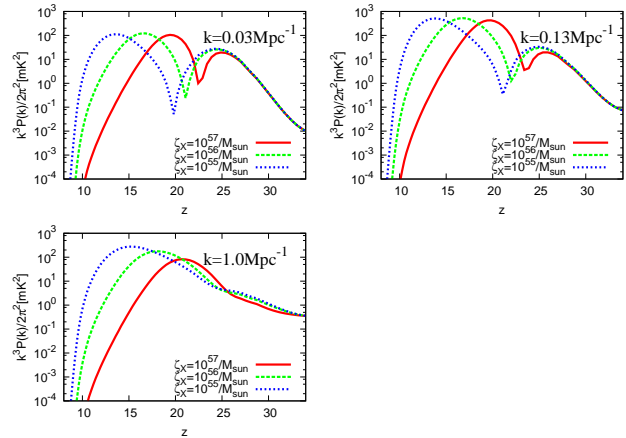}
\caption{Power spectra of the brightness temperature contributed from the fluctuations in $1 - T_\gamma / T_{\rm S}$ for $\zeta_X = 10^{57}~M_{\odot}^{-1}$ (red),  $10^{56}~M_{\odot}^{-1}$ (green) and $10^{55}~M_{\odot}^{-1}$ (blue) at three wave numbers as functions of redshift.}
\label{fig:ps_Ts_comp}
\end{figure}

Finally, we vary the number of X-ray photons emitted per solar mass, $\zeta_X$ \cite{Mesinger:2012ys}, and see the one-point statistics again. We take $\zeta_X = 10^{55}~M_{\odot}^{-1}, 10^{56}~M_{\odot}^{-1}$ (fiducial value), and $10^{57}~M_{\odot}^{-1}$, fixing the other parameters. The left top of Fig. \ref{fig:model} is the thermal history and we see that the spin temperature rises earlier for larger $\zeta_X$. The other panels of Fig. \ref{fig:model} show the evolution of the variance and skewness of the PDF of $\eta$, and the cross correlation between matter and $\eta$ fluctuations. As $\zeta_X$ increases, the critical redshift where the variance has a local minimum and the sign of the skewness and cross correlation changes increases. This is the expected behavior from our interpretation given in the previous section because X-ray heating will become effective earlier for larger $\zeta_X$. The power spectra of the brightness temperature contributed from the fluctuations in $\eta$ are plotted in Fig. \ref{fig:ps_Ts_comp} and, as expected again, the dip appears earlier for larger $\zeta_X$.

In summary, the dip in the evolution of power spectrum contributed from the spin temperature fluctuations can be understood as the state that X-ray heating is effective. The transition occurs at $z \approx 23$ for a fiducial set of model parameters and this depends on the effectiveness of X-ray heating. Conversely, our interpretation implies that, if we detect a dip in the redshift dependence of the power spectrum at the relatively higher redshift, we could know the redshift when X-ray begins to become effective. Further, when we detect a peak, it would be possible to know whether X-ray heating is effective at the redshift from the skewness. However, although the spin-temperature fluctuations are dominant at large scales, the contribution from matter fluctuations is not negligible and change the critical redshift as we see in the next section.

Here, it is important to note the difference between the redshift of the local minimum of the average spin temperature and the above critical redshift. The difference is due to the fact that the average spin temperature is a global property of the intergalactic gas while the skewness is largely affected by the behavior of a small fraction of gas near stars. Thus, the latter is more sensitive to the onset of X-ray heating. If so, it is expected that the difference in the two redshifts depends on the spectrum of X-ray \cite{Fialkov:2014kta}. For example, if the X-ray spectrum is hard, the difference will become smaller because higher-energy X-rays have larger mean free path so that they tend to heat larger region around the source.

\section{One-point statistics of brightness temperature}

In this section, we focus on the one point statistics of the brightness temperature. The variance and skewness of brightness temperature are described by 

\begin{align}
\sigma_{\delta T_{b}} &= (\overline{\delta T_{b}})^{2}[\sigma_{\delta_{m}}+\sigma_{\delta_{\eta}}+\sigma_{\delta_{{\rm x_{{\rm H}}}}} \nonumber \\
&\quad +\langle \delta_{m}\delta_{\eta} \rangle+\langle \delta_{m}\delta_{{\rm x_{{\rm H}}}} \rangle+\langle \delta_{\eta}\delta_{{\rm x_{{\rm H}}}}\rangle+O(\delta^{3})].
\label{eq:variance_dT}
\end{align}

\begin{align}
\gamma_{\delta T_{b}} &=(\overline{\delta T_{b}})^{3}[\gamma_{\delta_{m}}+\gamma_{\delta_{\eta}}+\gamma_{\delta_{x_{{\rm H}}}}+\langle \delta_{m}\delta_{\eta}\delta_{{\rm x_{{\rm H}}}}\rangle \nonumber \\
&\quad +3(\langle \delta^{2}_{m}\delta_{\eta} \rangle+\langle \delta^{2}_{m}\delta_{{\rm x_{{\rm H}}}} \rangle+\langle \delta^{2}_{\eta}\delta_{{\rm x_{{\rm H}}}}\rangle\nonumber \\
&\qquad+\langle \delta_{m}\delta^{2}_{\eta} \rangle+\langle \delta_{m}\delta^{2}_{{\rm x_{{\rm H}}}} \rangle+\langle \delta_{\eta}\delta^{2}_{{\rm x_{{\rm H}}}}\rangle)+O(\delta^{4})].
\label{eq:skewness_dT}
\end{align}

Here, we notice that $\overline{\delta T_{b}}$ is slightly different from the average $\langle \delta T \rangle$ obtained from brightness temperature map due to the contribution of higher-order terms. Although we focus on relatively high redshift, we find that $\delta_{{\rm x_{{\rm H}}}}$ is important when we consider the skewness.

\begin{figure}
\centering
\includegraphics[width=1.3\hsize, bb=0 0 380 290]{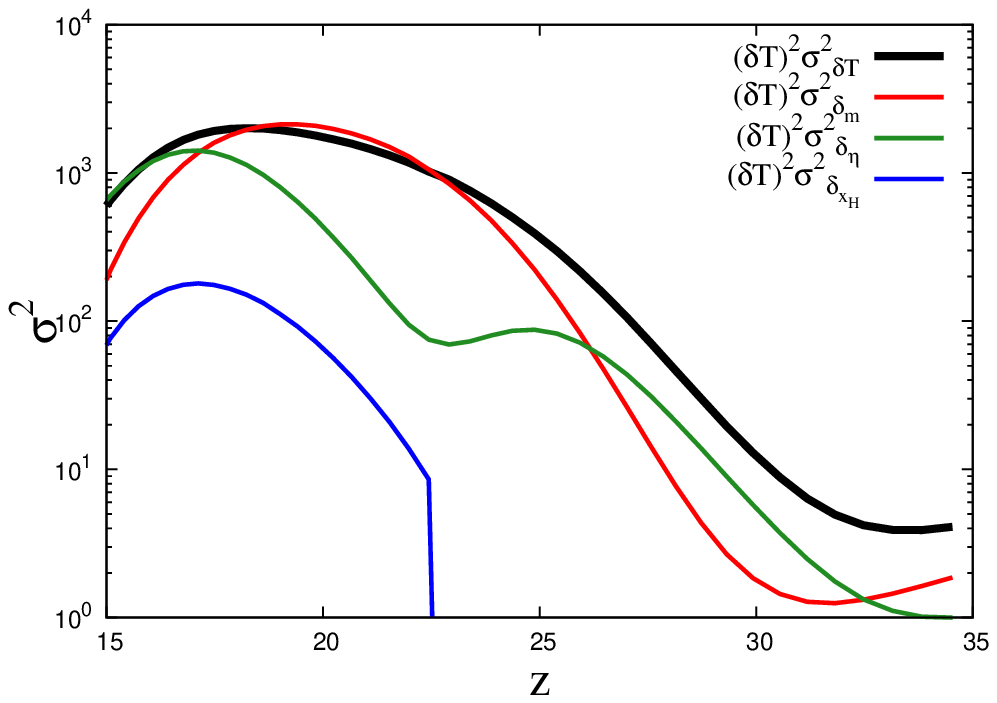}
\includegraphics[width=1.3\hsize, bb=0 0 380 290]{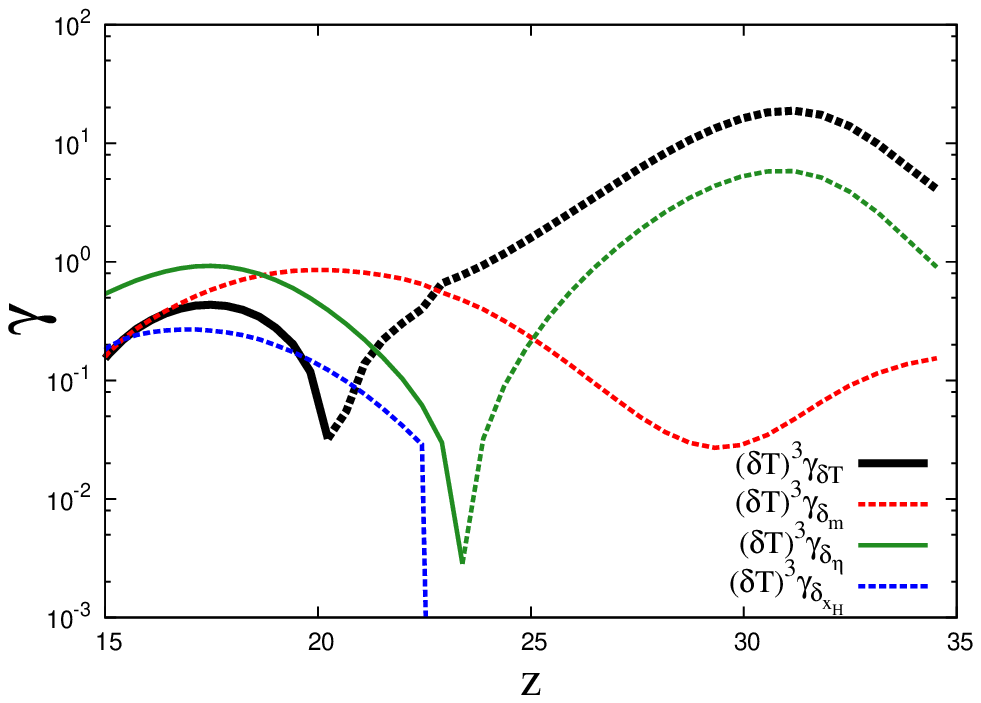}
\caption{[Top] The variance of brightness temperature(black), of $\delta_{m}$(red), of $\delta_{\eta}$(green), of $\delta_{{\rm x_{{\rm H}}}}$(blue).[Bottom] the skewness of brightness temperature(black), of $\delta_{m}$(red), of $\delta_{\eta}$(green), of $\delta_{{\rm x_{{\rm H}}}}$(blue). Dashed part corresponds to negative value and solid part corresponds to positive value.  All quantities are multiplied $(\overline{\delta T_{b}})^{2}$ (variance) or $(\overline{\delta T_{b}})^{3}$(skewness).}
\label{fig:skewness_dT2}
\end{figure}

In Fig.\ref{fig:skewness_dT2}, we plot the variance and skewness of brightness temperature with their components. Comparing the variance with skewness, we find that neutral-fraction fluctuation $\delta_{x_{\rm H}}$ is not negligible in skewness at $z \le 20$, although it does not contribute to the total variance so much. As you can see in the bottom of Fig. \ref{fig:skewness_dT2}, the change of sign in skewness for the brightness temperature is deviated from the one of $\eta$ due to the contribution of matter fluctuations, which have always negative skewness. Nontheless, the skewness is still a good indicator to study the epoch when X-ray heating becomes effective. We note here that the non-linear terms in skewness is the same order as the linear terms but qualitative behavior does not change whether we include non-linear terms or not.

\section{Summary \& Discussion}

In this paper, we gave a physical interpretation of the evolution of the power spectrum of the 21cm brightness temperature during the Cosmic Dawn and the Epoch of Reionization using a public code, 21cmFAST. With a fixed wave number, the power spectrum has three peaks as a function of redshift. First, we decomposed the power spectrum into those contributed from the fluctuations in dark matter density, spin temperature and neutral fraction and found that the peak with the lowest redshift is mostly contributed from the neutral fraction, while the other two peaks are contributed from the dark matter density and spin temperature fluctuations. Further, it was found that a dip between two peaks with higher redshifts is induced by the neutral fraction. Then, to understand the physical meaning of the dip, we investigated the one-point function of the spin temperature distribution. We found that the redshift of the dip is critical in a sense that the skewness of the one-point function and the correlation coefficient between the spin temperature and dark matter distribution change their signs and that the variance also has a dip. From this fact, it was implied that the dip in the power spectrum of the brightness temperature contributed from the spin temperature is a signature of the onset of X-ray heating of the gas. This interpretation was justified by seeing the behavior when varying the model parameter corresponding to X-ray heating.

Due to the contribution of dark matter density, the redshift of the dip in the power spectrum contributed from the spin temperature is slightly different from that in the full power spectrum. However, the dip will still be a good indicator of the onset of X-ray heating. Further, based on the cosmological perturbation theory, we can theoretically estimate the skewness of the brightness temperature contributed from the dark matter fluctuations in the standard cosmological model. Therefore, in principle, we can subtract the dark matter contribution and directly measure the skewness of the contribution from the spin temperature. Moreover, the scale dependence of the bispectrum, which has a close relation to the skewness, of the brightness temperature would help us to extract the information about the spin temperature from the observed brightness temperature.  The bispectrum is also deeply related to the non-linear terms \cite{Lidz:2006vj} in the power spectrum which we neglected in this paper. Because the fluctuations in the spin temperature are of order $\sim O(0.1)$ and are deviated from Gaussian distribution, the non-linear terms are potentially important. In particular, we found that the linear and non-linear terms are comparable when we evaluate the skewness. Further, the bispectrum and higher-order statistics would be very useful to extract physical information from 21cm signal \cite{Pillepich:2006fj,Cooray:2004kt,Cooray:2008eb}, as mentioned above. These topics will be presented elsewhere.

The detectability of variance and skewness was discussed in \cite{Harker:2008pz, 2008MNRAS.389.1319J, Watkinson:2013fea}.  
Following \cite{Watkinson:2013fea}, 
the signal-to-noise ratio, S/N, of skewness can be roughly estimated by
\begin{equation}
S/N \sim \sqrt{\frac{\gamma^2 \sigma^3}{\sigma_{\rm noise}^{6}/N_{\rm pix}}},
\label{eq:sn}
\end{equation}where we have neglected the contribution from kurtosis.
Here, $N_{\rm pix}$ is the number of pixels and we set $N_{\rm pix}=300^{3}$ and $\sigma_{\rm noise}$ is the instrumental noise on brightness temperature given by
\begin{eqnarray}
\sigma_{\rm noise}
&=& 0.37 {\rm mK} \left(\frac{10^6 {\rm m}}{A_{{\rm tot}}}\right)
                  \left(\frac{5^{'}}{\Delta \theta}\right)^{2}
                  \left(\frac{1+z}{10}\right)^{4.6} \nonumber \\
& & \times \sqrt{\left(\frac{1~{\rm MHz}}{\Delta \nu}
                       \frac{1000~{\rm hours}}{t_{\rm int}}\right)},
\label{eq:noise}
\end{eqnarray}where $A_{\rm tot}$ is the total effective area of array, $\theta$ is angular resolution, $\Delta \nu$ is frequency resolution and $t_{\rm int}$ is observing time. Considering the SKA1 ($A_{\rm tot}= 4 \times10^5 [{\rm m}^2]$) and SKA2 ($A_{\rm tot}= 3 \times10^6 [{\rm m}^2]$) with 1000 hours of observation time and $\Delta \nu \sim 0.1 {\rm MHz}$, we obtain S/N = 6 and 18 at $z = 15$, and S/N = 2 and 8 at $z = 20$, for SKA1 and SKA2, respectively.

%\begin{table*}
%\centering
%\caption{Signal to noise ratio of skewness at $z$=15, 20 for SKA1 and SKA2 observations.}
%\scalebox{1.5}{
%	\begin{tabular}{|l||l|c|l|l|} \hline
%	Observation & $A_{{\rm tot}}$ $[{\rm m}^{2}]$ & angle resolution & S/N(z=15) & S/N(z=20) \\ \hline \hline
%	SKA1  & 4 $\times$ $10^{5}$ & $\sim$ 3 arcmin & $\sim$ O(1) & $\sim$ O(0.1) \\ \hline
%	SKA2  & 3.2 $\times$ $10^{6}$ & $\sim$1 arcmin & $\sim$ O(10) & $\sim$  O(1) \\ \hline
%	\end{tabular}
%	}
%	\label{tb:sn}
%\end{table*}

\begin{figure}
\centering
\includegraphics[width=1.3\hsize, bb=0 0 380 290]{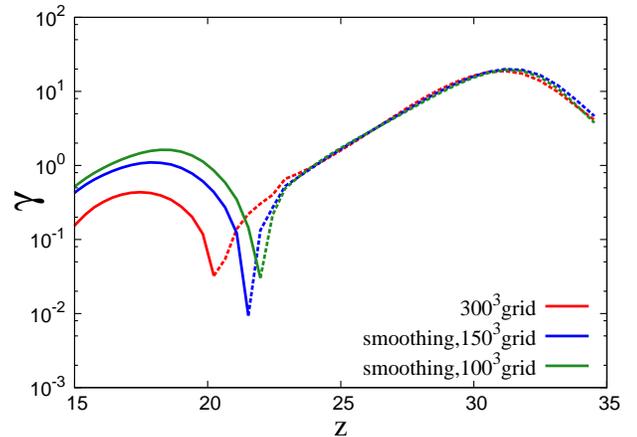}
\caption{Skewness evolution of the brightness temperature varying the number of grids: $300^3$ (fiducial, red), $150^3$ (blue) and $100^3$ (green). The latter two cases are calculated by smoothing the fiducial model.}
\label{fig:skewness_grid}
\end{figure}

Finally, we would like to note that the variance and skewness are actually dependent on the angular resolution and survey area. 
This can be understood by the fact that they are expressed by the integration of power spectrum and bispectrum with respect to the wavenumber. 
In our calculation, we have fixed the box size and the number of grids in simulations to be $(200~{\rm Mpc})^3$ and $300^3$, respectively.
This is corresponding to $0.66~{\rm Mpc}$ resolution or $\sim 3~{\rm arcmin}$ at $80~{\rm MHz}$ ($z = 17$).
In Fig. \ref{fig:skewness_grid}, we show the evolution of skewness of brightness temperature with varying the spacial resolution $0.66, 1.3$ and $2.0~{\rm Mpc}$. These correspond to 3 arcmin, 6 arcmin, 9 arcmin. and fixing the box size. The lowest resolution roughly corresponds to that of the SKA2. It is seen that the peak at $z \sim 18$ is diminished for higher resolution, while the one at $z \sim 30$ is not affected. On the other hand, the change of sign is shifted slightly to higher redshift for higher resolution. More detailed study on the dependence of skewness on the resolution needs the understanding of the bispectrum of the brightness temperature fluctuations and this will be presented elsewhere.

%\begin{table*}
%\centering
%\caption{Resolution at 80 MHz ($z \sim 17$)}
%\scalebox{1.5}{
%	\begin{tabular}{|l|c|r|r|} \hline
%	telescope & maximum baseline & spacial resolution & angle resolution\\ \hline \hline
%	LOFAR & $\sim$ 1500 m & $\sim$ 13 Mpc  & $\sim$ 4 arcmin\\ \hline
%	MWA   & $\sim$  750 m & $\sim$ 27 Mpc & $\sim$ 9 arcmin \\ \hline
%	SKA1  & $\sim$ 2000 m & $\sim$ 10 Mpc & $\sim$ 3 arcmin \\ \hline
%	SKA2  & $\sim$ 5000 m & $\sim$  4 Mpc & $\sim$1 arcmin\\ \hline
%	\end{tabular}
%	}
%	\label{tb:spec}
%\end{table*}

\section*{Acknowledgement}
We would like to thank K. Hasegawa for useful comments. This work is supported by Grant-in-Aid from the Ministry of Education, Culture, Sports, Science and Technology (MEXT) of Japan, Nos. 24340048(K.T. and K.I.) and 26610048. (K.T.), No. 25-3015(H.S.) and No. 24-2775(S.Y.).

\label{lastpage}


\begin{thebibliography}{99}

  %\cite{Ade:2013ydc}
\bibitem[Ade et al. 2013]{Ade:2013ydc} 
  P.~A.~R.~Ade {\it et al.}  [Planck Collaboration],
  %``Planck 2013 Results. XXIV. Constraints on primordial non-Gaussianity,''
  arXiv:1303.5084 [astro-ph.CO].
  %%CITATION = ARXIV:1303.5084;%%
  %384 citations counted in INSPIRE as of 20 Oct 2014
  
  
   %\cite{Baek:2010cm}
\bibitem[Baek et al. 2010]{Baek:2010cm}
  S.~Baek, B.~Semelin, P.~Di Matteo, Y.~Revaz and F.~Combes,
  %``Reionization by UV or X-ray sources,''
  arXiv:1003.0834 [astro-ph.CO].
  %%CITATION = ARXIV:1003.0834;%%
  
 \bibitem[Barkana \& Loeb 2008]{2008MNRAS.384.1069B}
R.~Barkana, \& A.~Loeb,\ 2008, MNRAS, 384, 1069   

 %\cite{Bartolo:2004if}
\bibitem[Bartolo et al. 2004]{Bartolo:2004if} 
  N.~Bartolo, E.~Komatsu, S.~Matarrese and A.~Riotto,
  %``Non-Gaussianity from inflation: Theory and observations,''
  Phys.\ Rept.\  {\bf 402}, 103 (2004)
  [astro-ph/0406398].
  %%CITATION = ASTRO-PH/0406398;%%
  %694 citations counted in INSPIRE as of 20 Oct 2014
  
  %\cite{Carilli:2014vha}
\bibitem[Carilli 2014]{Carilli:2014vha}
  C.~L.~Carilli,
  %``Square Kilometre Array key science: a progressive retrospective,''
  arXiv:1408.5317 [astro-ph.IM].
  %%CITATION = ARXIV:1408.5317;%%
  
     %\cite{Christian:2013gma}
\bibitem[Christian et al. 2013]{Christian:2013gma}
  P.~Christian and A.~Loeb,
  %``Measuring the X-ray background in the reionization era with first generation 21 cm experiments,''
  JCAP {\bf 1309} (2013) 014
  [arXiv:1305.5541 [astro-ph.CO]].
  %%CITATION = ARXIV:1305.5541;%%
  %1 citations counted in INSPIRE as of 17 Sep 2014
  
    %\cite{Cooray:2004kt}
\bibitem[Cooray 2005]{Cooray:2004kt}
  A.~Cooray,
  %``Large-scale non-Gaussianities in the 21 cm background anisotropies from the era of reionization,''
  Mon.\ Not.\ Roy.\ Astron.\ Soc.\  {\bf 363} (2005) 1049
  [astro-ph/0411430].
  %%CITATION = ASTRO-PH/0411430;%%
  %12 citations counted in INSPIRE as of 24 Sep 2014
  
  %\cite{Cooray:2008eb}
\bibitem[Cooray et al. 2008]{Cooray:2008eb}
  A.~Cooray, C.~Li and A.~Melchiorri,
  %``The trispectrum of 21-cm background anisotropies as a probe of primordial non-Gaussianity,''
  Phys.\ Rev.\ D {\bf 77} (2008) 103506
  [arXiv:0801.3463 [astro-ph]].
  %%CITATION = ARXIV:0801.3463;%%
  %25 citations counted in INSPIRE as of 24 Sep 2014
  
 \bibitem[Dewdney et al 2013]{Dewdney:2013}
 P. Dewdney et al
 $http://www.skatelescope.org/wp-content/uploads/2012/07/SKA-TEL-SKO-DD-001-1_BaselineDesign1.pdf$
 
  \bibitem[Dillon et al 2013]{Dillon:2013rfa} 
  J.~S.~Dillon, A.~Liu, C.~L.~Williams, J.~N.~Hewitt, M.~Tegmark, E.~H.~Morgan, A.~M.~Levine and M.~F.~Morales {\it et al.},
  %``Overcoming real-world obstacles in 21 cm power spectrum estimation: A method demonstration and results from early Murchison Widefield Array data,''
  Phys.\ Rev.\ D {\bf 89}, 023002 (2014)
  [arXiv:1304.4229 [astro-ph.CO]].
  %%CITATION = ARXIV:1304.4229;%%
  %13 citations counted in INSPIRE as of 10 Nov 2014
  
    %\cite{Fan:2006dp}
\bibitem[Fan et al. 2006]{Fan:2006dp}
  X.~H.~Fan, C.~L.~Carilli and B.~G.~Keating,
  %``Observational constraints on cosmic reionization,''
  Ann.\ Rev.\ Astron.\ Astrophys.\  {\bf 44} (2006) 415
  [astro-ph/0602375].
  %%CITATION = ASTRO-PH/0602375;%%
  %268 citations counted in INSPIRE as of 17 Sep 2014

 %\cite{Fialkov:2014kta}
 
   %\cite{Fialkov:2013uwm}
\bibitem[Fialkov et al. 2013]{Fialkov:2013uwm} 
  A.~Fialkov, R.~Barkana, A.~Pinhas and E.~Visbal,
  %``Complete history of the observable 21-cm signal from the first stars during the pre-reionization era,''
  arXiv:1306.2354 [astro-ph.CO].
  %%CITATION = ARXIV:1306.2354;%%
  %2 citations counted in INSPIRE as of 17 Oct 2014
 
\bibitem[Fialkov et al 2014]{Fialkov:2014kta} 
  A.~Fialkov, R.~Barkana and E.~Visbal,
  %``The observable signature of late heating of the universe during cosmic reionization,''
  arXiv:1402.0940 [astro-ph.CO].
  %%CITATION = ARXIV:1402.0940;%%
  
  %\cite{Furlanetto:2006jb}
\bibitem[Furlanetto et al 2006]{fur}
S. Furlanetto, S. P. Oh and F. Briggs,
  %``Cosmology at Low Frequencies: The 21 cm Transition and the High-Redshift universe,''
  Phys.\ Rept.\  {\bf 433} (2006) 181
  [astro-ph/0608032].
  %%CITATION = ASTRO-PH/0608032;%%
  %312 citations counted in INSPIRE as of 17 Sep 2014
  
  %\cite{Ghara:2014yfa}
\bibitem[Ghara et al. 2014]{Ghara:2014yfa}
  R. Ghara, T. R. Choudhury and K. K. Datta,
  %``21 cm signal from cosmic dawn: Imprints of spin temperature fluctuations and peculiar velocities,''
  arXiv:1406.4157 [astro-ph.CO].
  %%CITATION = ARXIV:1406.4157;%%
  %1 citations counted in INSPIRE as of 12 Aug 2014
  %\cite{Lidz:2006vj}
  
    %\cite{Harker:2008pz}
\bibitem[Harker et al. 2008]{Harker:2008pz} 
  G.~J.~A.~Harker, S.~Zaroubi, R.~M.~Thomas, V.~Jelic, P.~Labropoulos, G.~Mellema, I.~T.~Iliev and G.~Bernardi {\it et al.},
  %``Detection and extraction of signals from the epoch of reionization using higher order one-point statistics,''
  arXiv:0809.2428 [astro-ph].
  %%CITATION = ARXIV:0809.2428;%%
  %2 citations counted in INSPIRE as of 18 Aug 2014
  
  \bibitem[Jeli{\'c} et al.2008]{2008MNRAS.389.1319J} Jeli{\'c}, V., 
Zaroubi, S., Labropoulos, P., et al.\ 2008, MNRAS, 389, 1319 
  
   %\cite{Komatsu:2010fb}
\bibitem[Komatsu et al. 2011]{Komatsu:2010fb}
  E.~Komatsu {\it et al.}  [WMAP Collaboration],
  %``Seven-Year Wilkinson Microwave Anisotropy Probe (WMAP) Observations: Cosmological Interpretation,''
  Astrophys.\ J.\ Suppl.\  {\bf 192} (2011) 18
  [arXiv:1001.4538 [astro-ph.CO]].
  %%CITATION = ARXIV:1001.4538;%%
  %4726 citations counted in INSPIRE as of 17 Sep 2014
  
  
  \bibitem[Litz et al. 2007]{Lidz:2006vj}
  A.~Lidz, O.~Zahn, M.~McQuinn, M.~Zaldarriaga and S.~Dutta,
  %``Higher Order Contributions to the 21 cm Power Spectrum,''
  Astrophys.\ J.\  {\bf 659} (2007) 865
  [astro-ph/0610054].
  %%CITATION = ASTRO-PH/0610054;%%
  %29 citations counted in INSPIRE as of 20 Sep 2014
  
    %\cite{Loeb:2000fc}
\bibitem[Loeb \& Barkana 2001]{Loeb:2000fc} 
  A.~Loeb and R.~Barkana,
  %``The Reionization of the universe by the first stars and quasars,''
  Ann.\ Rev.\ Astron.\ Astrophys.\  {\bf 39}, 19 (2001)
  [astro-ph/0010467].
  %%CITATION = ASTRO-PH/0010467;%%
  %154 citations counted in INSPIRE as of 19 Oct 2014
  
  \bibitem[Loeb et al 2008]{Loeb:2008hg} 
  A.~Loeb and S.~Wyithe,
  %``Precise Measurement of the Cosmological Power Spectrum With a Dedicated 21cm Survey After Reionization,''
  Phys.\ Rev.\ Lett.\  {\bf 100}, 161301 (2008)
  [arXiv:0801.1677 [astro-ph]].
  %%CITATION = ARXIV:0801.1677;%%
  %36 citations counted in INSPIRE as of 10 Nov 2014
  
   %\cite{Mao:2008ug}
\bibitem[Mao et al. 2008]{Mao:2008ug} 
  Y.~Mao, M.~Tegmark, M.~McQuinn, M.~Zaldarriaga and O.~Zahn,
  %``How accurately can 21 cm tomography constrain cosmology?,''
  Phys.\ Rev.\ D {\bf 78}, 023529 (2008)
  [arXiv:0802.1710 [astro-ph]].
  %%CITATION = ARXIV:0802.1710;%%
  %92 citations counted in INSPIRE as of 10 Nov 2014
  
 \bibitem[Mellema et al. 2013]{2013ExA....36..235M} 
 G. Mellema,  L.~V.~E. Koopmans, F.~A.Abdalla. et al\ 2013, Experimental Astronomy, 36, 235 
  
  %\cite{Mesinger:2007pd}
\bibitem[Mesinger et al. 2007]{Mesinger:2007pd} 
  A.~Mesinger and S.~Furlanetto,
  %``Efficient Simulations of Early Structure Formation and Reionization,''
  arXiv:0704.0946 [astro-ph].
  %%CITATION = ARXIV:0704.0946;%%
  %12 citations counted in INSPIRE as of 09 Oct 2014
  
  \bibitem[Mesinger et al. 2011]{2011MNRAS.411..955M} 
A.~Mesinger, S.~Furlanetto, R.~Cen,2011, MNRAS, 411, 955 
  
     %\cite{Mesinger:2012ys}
\bibitem[Mesinger et al 2012]{Mesinger:2012ys}
  A.~Mesinger, A.~Ferrara and D.~S.~Spiegel,
  %``Signatures of X-rays in the early universe,''
  arXiv:1210.7319 [astro-ph.CO].
  %%CITATION = ARXIV:1210.7319;%%
  %4 citations counted in INSPIRE as of 17 Sep 2014
  
    %\cite{Mesinger:2013nua}
\bibitem[Mesinger et al. 2013]{Mesinger:2013nua}
  A.~Mesinger, A.~Ewall-Wice and J.~Hewitt,
  %``Reionization and Beyond: detecting the peaks of the cosmological 21cm signal,''
  arXiv:1310.0465 [astro-ph.CO].
  %%CITATION = ARXIV:1310.0465;%%
  %4 citations counted in INSPIRE as of 17 Sep 2014
  
        
  %\cite{Pillepich:2006fj}
\bibitem[Pillepich et al. 2006]{Pillepich:2006fj}
  A.~Pillepich, C.~Porciani and S.~Matarrese,
  %``The bispectrum of redshifted 21-cm fluctuations from the dark ages,''
  Astrophys.\ J.\  {\bf 662} (2007) 1
  [astro-ph/0611126].
  %%CITATION = ASTRO-PH/0611126;%%
  %54 citations counted in INSPIRE as of 24 Sep 2014
  
    %\cite{Pober:2014aca}
\bibitem[Pober et al a. 2014]{Pober:2014aca}
  D.~C.~J.~J.~C.~Pober, A.~R.~Parsons, J.~E.~Aguirre, Z.~Ali, J.~Bowman, R.~F.~Bradley, C.~L.~Carilli and D.~R.~DeBoer {\it et al.},
  %``Multi-redshift limits on the 21cm power spectrum from PAPER,''
  arXiv:1408.3389 [astro-ph.CO].
  %%CITATION = ARXIV:1408.3389;%%
  %1 citations counted in INSPIRE as of 17 Sep 2014
  
  \bibitem[Pober et al b. 2014]{2014ApJ...782...66P}
J.~C.~Pobe, A.~Liu, A, J.~S.~Dillon, et al.\ 2014, ApJ, 782, 66 

  
 %\cite{Pritchard:2006sq}
\bibitem[Pritchard \& Furlanetto 2007]{Pritchard:2006sq}
  J.~R.~Pritchard and S.~R.~Furlanetto,
  %``21 cm fluctuations from inhomogeneous X-ray heating before reionization,''
  Mon.\ Not.\ Roy.\ Astron.\ Soc.\  {\bf 376} (2007) 1680
  [astro-ph/0607234].
  %%CITATION = ASTRO-PH/0607234;%%
  %82 citations counted in INSPIRE as of 17 Sep 2014
  
    %\cite{Rottgering:2003jh}
\bibitem[Rottgering et al. 2013]{Rottgering:2003jh}
  H.~Rottgering,
  %``LOFAR, A New low frequency radio telescope,''
  New Astron.\ Rev.\  {\bf 47} (2003) 405
  [astro-ph/0309537].
  %%CITATION = ASTRO-PH/0309537;%%
  %38 citations counted in INSPIRE as of 17 Sep 2014
  
  \bibitem[Santos et al. 2008]{2008ApJ...689....1S}
M.~G.~Santos, A.~Amblard, J.~Pritchard et al.\ 2008, ApJ, 689, 1 

\bibitem[Santos et al. 2011]{2011A&A...527A..93S} M.~G.~Santos, M.~B.~Silva, J.~R.~Pritchard, R.~Cen,\& A.~Cooray.\ 2011, A \& A, 527, A93

  
    %\cite{Tingay:2012ps}
\bibitem[Tingay et al. 2012]{Tingay:2012ps}
  S.~J.~Tingay, R.~Goeke, J.~D.~Bowman, D.~Emrich, S.~M.~Ord, D.~A.~Mitchell, M.~F.~Morales and T.~Booler {\it et al.},
  %``The Murchison Widefield Array: the Square Kilometre Array Precursor at low radio frequencies,''
  arXiv:1206.6945 [astro-ph.IM].
  %%CITATION = ARXIV:1206.6945;%%
  %26 citations counted in INSPIRE as of 17 Sep 2014
  
  
  
    %\cite{Visbal:2012aw}
\bibitem[Visbal et al. 2012]{Visbal:2012aw} 
  E.~Visbal, R.~Barkana, A.~Fialkov, D.~Tseliakhovich and C.~Hirata,
  %``The signature of the first stars in atomic hydrogen at redshift 20,''
  arXiv:1201.1005 [astro-ph.CO].
  %%CITATION = ARXIV:1201.1005;%%
  %7 citations counted in INSPIRE as of 17 Oct 2014
  
    %\cite{Watkinson:2013fea}
\bibitem[Watkinson \& Pritchard 2013]{Watkinson:2013fea} 
  C.~A.~Watkinson and J.~R.~Pritchard,
  %``Distinguishing models of reionization using future radio observations of 21-cm one-point statistics,''
  Mon.\ Not.\ Roy.\ Astron.\ Soc.\  {\bf 443}, 3090 (2014)
  [arXiv:1312.1342 [astro-ph.CO]].
  %%CITATION = ARXIV:1312.1342;%%
  %1 citations counted in INSPIRE as of 18 Aug 2014

\bibitem[Wouthuysen 1952]{wou}
S.~A.~Wouthuysen, Astronomical Journal, 57,31


\bibitem[Yoshida et al. 2006]{yos}
N.~Yoshida, K.~Omukai, L.~Hernquist, \& T.~Abel \ 2006, ApJ, 652, 6




  

  
 
    

  
 
 

      
  
  

  

  

  
  

   

      

    

   







    

    
\end{thebibliography}
\end{document}